\documentclass[osajnl,twocolumn,showpacs,superscriptaddress,10pt]{revtex4-1} 
\usepackage{amsmath,amssymb,graphicx}

\usepackage{amssymb,amsmath}
\usepackage{lmodern}
\usepackage{bm}
\usepackage{sansmath}
\usepackage{mathrsfs}

\newcommand{\degree}{\ensuremath{^\circ}}
\newcommand{\lambdaaerzero}{$\Lambda^0_{\mathrm{aer}}$~}

\newcommand{\lambdaaer}{$\Lambda_{\mathrm{aer}}$~}
\newcommand{\lambdamol}{$\Lambda_{\mathrm{mol}}$}

\newcommand{\haerzero}{$H^{0}_{\mathrm{aer}}$}
\newcommand{\rdir}{$r_{\mathrm{dir}}$~}
\newcommand{\rrel}{$r_{\mathrm{rel}}$~}
\newcommand{\thetarel}{$\theta_{\mathrm{rel}}$~}

\begin{document}

\title{Monte Carlo simulation of light scattering in the atmosphere and effect of atmospheric aerosols on the point spread function}

\author{Joshua Colombi}
\author{Karim Louedec}\email{Corresponding author: karim.louedec@lpsc.in2p3.fr}
\affiliation{Laboratoire de Physique Subatomique et de Cosmologie (LPSC), UJF-INPG, CNRS/IN2P3, 38026 Grenoble cedex, France}

\begin{abstract}
We present a Monte Carlo simulation for the scattering of light in the case of an isotropic light source. The scattering phase functions are studied particularly in detail to understand how they can affect the multiple light scattering in the atmosphere. We show that although aerosols are usually in lower density than molecules in the atmosphere, they can have a non-negligible effect on the atmospheric point spread function. This effect is especially expected for ground-based detectors when large aerosols are present in the atmosphere.
\end{abstract}


\maketitle

\section{Introduction}
Light coming from an isotropic source is scattered and/or absorbed by molecules and/or aerosols in the atmosphere. In the case of fog or rain, the single light scattering approximation -- when scattered light cannot be dispersed again to the detector and only direct light is recorded -- is not valid anymore. Thus, the multiple light scattering -- when photons are scattered several times before being detected -- has to be taken into account in the total signal recorded. Whereas the first phenomenon reduces the amount of light arriving at the detector, the latter increases the spatial blurring of the isotropic light source. Atmospheric blur occurs especially for long distances and total optical depths greater than unity. This effect is well known for light propagation in the atmosphere and has been studied by many authors. A nice review of relevant findings in this research field can be read in~\cite{Kopeika_3}. Originally, these studies began with satellites imaging Earth where aerosol blur is considered as the main source of atmospheric blur~\cite{Dave,Pearce,Kopeika_1,Kopeika_2}. This effect is usually called {\it the adjacency effect}~\cite{Otterman,Tanre,Reinersman} since photons scattered by aerosols are recorded in pixels adjacent to where they should be.

The problem of light scattering in the atmosphere has not analytical solutions. Even if analytical approximation solutions can be used in some cases~\cite{Kopeika_1,Ishimaru}, Monte Carlo simulations are usually used to study light propagation in the atmosphere. A multitude of Monte Carlo simulations have been developed in the past years, all yielding to similar conclusion: aerosol scattering is the main contribution to atmospheric blur, atmospheric turbulence being much less important. A significant source of atmospheric blur is especially aerosol scatter of light at near-forward angles~\cite{Kopeika_3,Reinersman}. The multiple scattering of light is affected by the optical thickness of the atmosphere, the aerosol size distribution and the aerosol vertical profile. Whereas many works have studied the effect of the optical thickness, the aerosol blur is also very dependent on the aerosol size distribution, and especially on the corresponding asymmetry parameter of the aerosol scattering phase function. The purpose of this work is to better explain the dependence of the aerosol blur on the aerosol size, and its corresponding effect on the atmospheric point spread function. Indeed, as explained previously, aerosol scattering at very forward angles is a significant source of blur and this phenomenon is strongly governed by the asymmetry parameter. Section~\ref{sec:modelling} is a brief introduction of some quantities concerning light scattering, before describing in detail the Monte Carlo simulation developed for this work. Section~\ref{sec:simu_results} gives a general overview of how scattered photons disperse across space for different atmospheric conditions. Then, in Section~\ref{sec:global_view}, we explain how different atmospheric conditions affect the multiple scattering contribution to the total light arriving at detectors within a given integration time across all space. This result is finally applied to the point spread function for a ground-based detector in Section~\ref{sec:ground_telescope}.

\section{Modelling and simulation of scattering in the atmosphere}
\label{sec:modelling}
Throughout this paper, the scatterers in the atmosphere will be modelled as non-absorbing spherical particles of different sizes~\cite{Hulst,Bohren}. Scatterers in the atmosphere are usually divided into two main types - aerosols and molecules.

\subsection{The density of scatterers in the atmosphere}
The attenuation length (or mean free path) $\Lambda$ associated with a given scatterer is related to its density and is the average distance that a photon travels before being scattered. For a given number of photons $N$ traveling across an infinitesimal distance d$l$, the amount scattered is given by $\mathrm{d}N^{\mathrm{scat}} = N\times\mathrm{d}l / \Lambda$. Density and $\Lambda$ are inversely related such that a higher value of $\Lambda$ is equivalent to a lower density of scatterers in the atmosphere. Molecules and aerosols have different associated densities in the atmosphere and are described respectively by a total attenuation length $\Lambda_{\mathrm{mol}}$ and $\Lambda_{\mathrm{aer}}$.  The value of these total attenuation lengths in the atmosphere can be modelled as horizontally uniform and exponentially increasing with respect to height above ground level $h_{\mathrm{agl}}$. The total attenuation length for each scatterer population is written as 
\begin{equation}
\begin{cases}
\Lambda_{\mathrm{mol}}(h_{\mathrm{agl}}) = \Lambda^{0}_{\mathrm{mol}} \,\exp \left[ (h_{\mathrm{agl}} + h_{\rm det})/H^{0}_\mathrm{mol} \right],\\
\Lambda_{\mathrm{aer}}(h_{\mathrm{agl}}) = \Lambda^{0}_{\mathrm{aer}} \, \exp\left[ h_{\mathrm{agl}}/ H^{0}_{\mathrm{aer}} \right],
\end{cases}
\label{eq:lambdas}
\end{equation}
where $\{\Lambda^{0}_{\mathrm{aer}}, \Lambda^{0}_{\mathrm{mol}}\}$ are multiplicative scale factors, $\{H^{0}_{\mathrm{aer}}, H^{0}_{\mathrm{mol}}\}$ are scale heights associated with aerosols and molecules, respectively, and $h_{\rm det}$ is the altitude difference between ground level and sea level. The \emph{US standard atmospheric model} is used to fix typical values for molecular component: $\Lambda^{0}_{\mathrm{mol}} = 14.2~$km and $H^{0}_{\mathrm{mol}} = 8.0~$km~\cite{Bucholtz}. These values are of course slightly variable with weather conditions~\cite{EPJP_BiancaMartin} but the effect of molecule concentration on multiply scattered light is not that of interest in this work. Atmospheric aerosols are found in lower densities than molecules in the atmosphere and are mostly present only in the first few kilometres above ground level. The aerosol population is much more variable in time than the molecular as their presence is dependent on many more factors such as the wind, rain and pollution~\cite{EPJP_Aerosol}. However, the model of the exponential distribution is usually used to describe aerosol populations. Only the parameter \lambdaaerzero will be varied and \haerzero~is fixed at $1.5~$km for the entirety of this work.

\subsection{The different scattering phase functions}
\label{sec:phase_functions}
A scattering phase function is used to describe the angular distribution of scattered photons. It is typically written as a normalised probability density function expressed in units of probability per unit of solid angle. When integrated over a given solid angle $\Omega$, a scattering phase function gives the probability of a photon being scattered with a direction that is within this solid angle range. Since scattering is always uniform in azimuthal angle $\phi$ for both aerosols and molecules, the scattering phase function is always written simply as a function of polar scattering angle $\psi$.

Molecules are governed by Rayleigh scattering which can be derived analytically via the approximation that the electromagnetic field of incident light is constant across the small size of the particle~\cite{Bucholtz}. The molecular phase function is written as
\begin{equation}
P_{\mathrm{mol}}(\psi)=\frac{3}{16 \pi}(1+\cos^2\psi),
\label{eq:RPF}
\end{equation}
where $\psi$ is the polar scattering angle and $P_{\mathrm{mol}}$ the probability per unit solid angle. The function $P_{\mathrm{mol}}$ is symmetric about the point $\pi/2$ and so the probability of a photon scattering in forward or backward directions is always equal for molecules.

\begin{figure*}[t!]
\centering
\includegraphics[width = 1.0\textwidth]{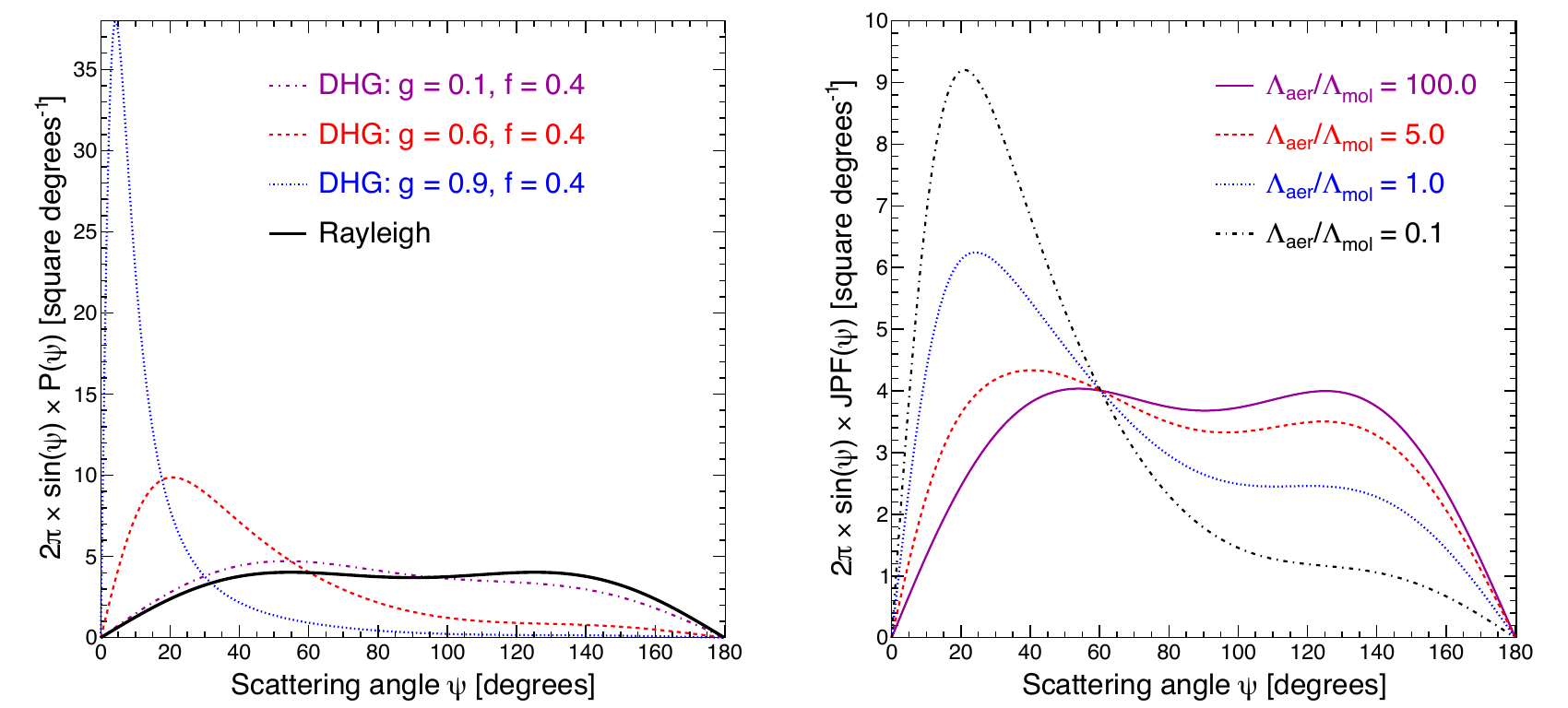}
\caption{(Color online) {\bf Scattering phase function per unit of polar angle $\psi$, and its dependence to atmospheric conditions.} Scattering phase functions are in units of probability per solid angle $\Omega$ as opposed to probability per unit of $\psi$ as necessary to get the probability density function of the polar angle $\psi$. Thus, the scattering phase functions $P_{\mathrm{mol}} (\psi)$ and $P_{\mathrm{aer}} (\psi)$ have to be multiplied by $2\pi \,\sin{\psi}$ to remove the solid angle weighting. (left) $ P_{\mathrm{aer}}(\psi)$ plotted for different values of $g$ and fixed $f=0.4$ used in the Double Henyey-Greenstein (DHG) phase function and $P_{\mathrm{mol}}(\psi)$ for the molecular phase function. (right) The joint probability phase function weighted by $2\pi\sin(\psi)$, with different ratios of $\Lambda_{\mathrm{aer}} / \Lambda_{\mathrm{mol}}$ ($g$ is kept equal to 0.6). }
\label{fig:APF}
\end{figure*}

Atmospheric aerosols typically come in the form of small particles of dust or droplets found in suspension in the atmosphere. The angular dependence of scattering by these particles is less easily described as the electromagnetic field of incident light can no longer be approximated as constant over the volume of the particle. Mie scattering theory~\cite{Mie} offers a solution in the form of an infinite series for the scattering of non-absorbing spherical objects of any size. The number of terms required in this infinite series to calculate the scattering phase function is given in~\cite{Wiscombe}, it is far too time consuming for the Monte Carlo simulations. As such, a parameterisation named the Double-Henyey Greenstein (DHG) phase function~\cite{HenyeyGreenstein,EPJP_Aerosol} is usually used. It is a parameterisation valid for various particle types and different media~\cite{HG_astro,HG_meteo,HG_bio}. It is written as
\begin{eqnarray}
P_{\mathrm{aer}}(\psi|g,f)= &\frac{1-g^2}{4\pi}\left[\frac{1}{(1+g^2-2g\cos{\psi})^\frac{3}{2}} +f\left(\frac{3\cos^2{\psi}-1}{2(1+g^2)^\frac{3}{2}}\right)\right]
\label{eq:APF}
\end{eqnarray}
where $g$ is the asymmetry parameter given by $\left<\cos\psi\right>$ and $f$ the backward scattering correction parameter. $g$ and $f$ vary in the intervals $[-1,1]$ and $[0 ,1]$, respectively. Most of the atmospheric conditions can be probed by varying the value of the asymmetry parameter $g$: aerosols ($0.2 \leq g \leq 0.7$), haze ($0.7 \leq g \leq 0.8$), mist ($0.8 \leq g \leq 0.85$), fog ($0.85 \leq g \leq 0.9$) or rain ($0.9 \leq g \leq 1.0$)~\cite{Metari}. Changing $g$ from 0.2 to 1.0 increases greatly the probability of scattering in the very forward direction as it can be observed in Fig.~\ref{fig:APF}(left). The reader is referred to ~\cite{Ramsauer_1,Ramsauer_2} to see the recently published work on the relation between $g$ and the mean radius of an aerosol: a physical interpretation of the asymmetry parameter $g$ in the DHG phase function is the mean aerosol size. The parameter $f$ is an extra parameter acting as a fine tune for the amount of backward scattering. It will be fixed at 0.4 for the rest of this work. 

A joint scattering phase function, weighting the aerosol and molecular phase functions by the corresponding densities of aerosols and molecules at a given position in space, gives the scattering phase function associated with any random scattering event. This joint scattering phase function can be written as
\begin{equation}
P_{\mathrm{jpf}}(\psi)= \frac{P_{\mathrm{aer}}(\psi)}{1 + \left(\frac{\Lambda_{\mathrm{aer}}}{\Lambda_{\mathrm{mol}}}\right)} +\frac{P_{\mathrm{mol}}(\psi)}{1 + \left( \frac{\Lambda_{\mathrm{mol}}}{\Lambda_{\mathrm{aer}}}\right)}.
\end{equation}
The use of this joint scattering phase function is relevant in understanding to what degree aerosols and molecules change the overall angular distribution of scattering at a given point in space. Figure~\ref{fig:APF}(right) displays the joint scattering phase function for different ratios of density of aerosols and molecules, all for a value of $g=0.6$ (typical value for aerosols in the atmosphere). It shows the probability of generating a scattering angle $\psi$ for different ratios of concentration of aerosols and molecules for a random scattering event that could be either caused by a molecule or aerosol. It is seen that as  \lambdaaer/\lambdamol~decreases ({\it i.e.}\ the density of aerosols increases), the high forward scattering peak associated with the aerosols becomes increasingly prominent.

\begin{figure*}[t!]
\centering
\includegraphics [width=1.0\textwidth] {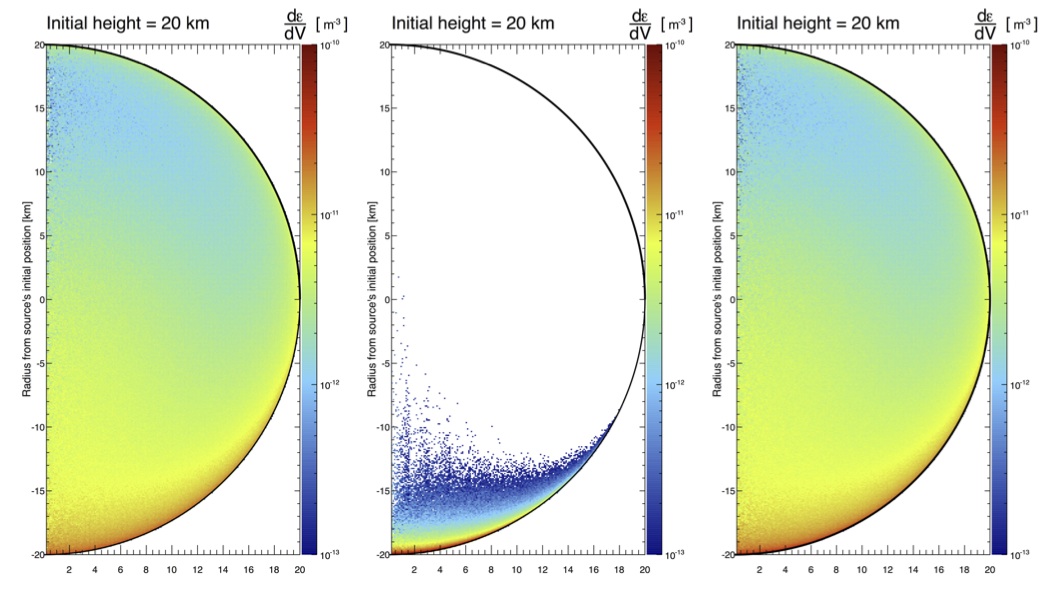}
\caption{(Color online) {\bf Relative effect on the density distribution of indirect photons for aerosols and molecules in a real atmosphere, illustrated by simulations with aerosols and molecules independently and simultaneously present.}  (left) An atmosphere consisting of molecules only. (middle) An atmosphere consisting of only aerosols with parameters $\{$g=0.6, \lambdaaerzero=10~km and \haerzero=1.5~km$\}$. (right) An atmosphere consisting simultaneously of both aerosols and molecules with the same atmospheric conditions.}
\label{fig:relativedistributions} 
\end{figure*}

\subsection{Monte Carlo code description}
\label{sec:code}
This Monte Carlo simulation code traces the paths of photons in the atmosphere between their isotropic source and a detector, accounting for changes in their vector position and their probability for having been attenuated. An isotropic light source is simulated by creating $N$ photons with the same initial position and isotropically distributed initial directions. To achieve isotropy in the initial directions, the azimuthal angle $\phi$ is generated randomly in the interval $[0, 2\pi[$ and the polar angle $\theta$ by $\theta = \cos^{-1}(\mathscr{R})$ with $\mathscr{R}$ randomly generated in the interval $[-1,1[$. The  formula $\theta = \cos^{-1}(\mathscr{R})$ accounts for the weighting  of the solid angle at a given $\theta$ equal to $\mathrm{d}(\cos\theta)$. Photons are then propagated through a given distance $D$. The step length d$l=c\times{\rm d}t$ used in the program is set as d$l=D/1000$. A photon is randomly scattered by an aerosol, molecule or not at all in accordance with the probabilities ${\rm d}l/{\Lambda_{\mathrm{aer}}}$, ${\rm d}l/{\Lambda_{\mathrm{mol}}}$ or $1 - {\rm d}l/{\Lambda_{\mathrm{aer}}} - {\rm d}l/{\Lambda_{\mathrm{mol}}}$, respectively. Scattering in the azimuthal angle $\phi$ is isotropic for both types. In contrast, as explained in the previous subsection, scattering in the polar angle $\psi$ is dependent on the scattering phase function involved: the Rayleigh and the Double Henyey-Greenstein scattering phase functions are used for molecular and aerosol scattering events, respectively. Finally, the polar coordinates relative to the source's initial position $\{r_{\mathrm{rel}}, \theta_{\mathrm{rel}}, \phi_{\mathrm{rel}} \}$ are used to store the position of all scattered photons at the end of each simulation.

A horizontally uniform density distribution for aerosols and molecules is assumed for the vertical profile of the atmosphere. Thus, using isotropy in azimuthal scattering angle $\phi$ for all scattering phase functions, a symmetry in the distribution of $\phi_{\mathrm{rel}}$ for all scattered photons should be found for a sufficiently high number of initial photons. The isotropy in $\phi_{\mathrm{rel}}$ means that any data found at a constant $\{r_{\mathrm{rel}} , \theta_{\mathrm{rel}}\}$ is the same for all values of $\phi_{\mathrm{rel}}[0,2\pi[$. Thus, all information given on the 3-D distribution in space can be given in terms of $r_{\mathrm{rel}}$ and $\theta_{\mathrm{rel}}$ only. The present work does not investigate the effect of a change of the vertical distribution of aerosols ({\it i.e.}\ exponential shape and vertical aerosol scale $H^{0}_{\mathrm{aer}}$), nor the effect of overlying cirrus clouds or aerosol layers on the multiple scattered light contribution to direct light. The next section presents a general overview of how scattered photons disperse across space for different atmospheric conditions.

\begin{figure*}[t!]
\centering
\includegraphics [width=1.0\textwidth] {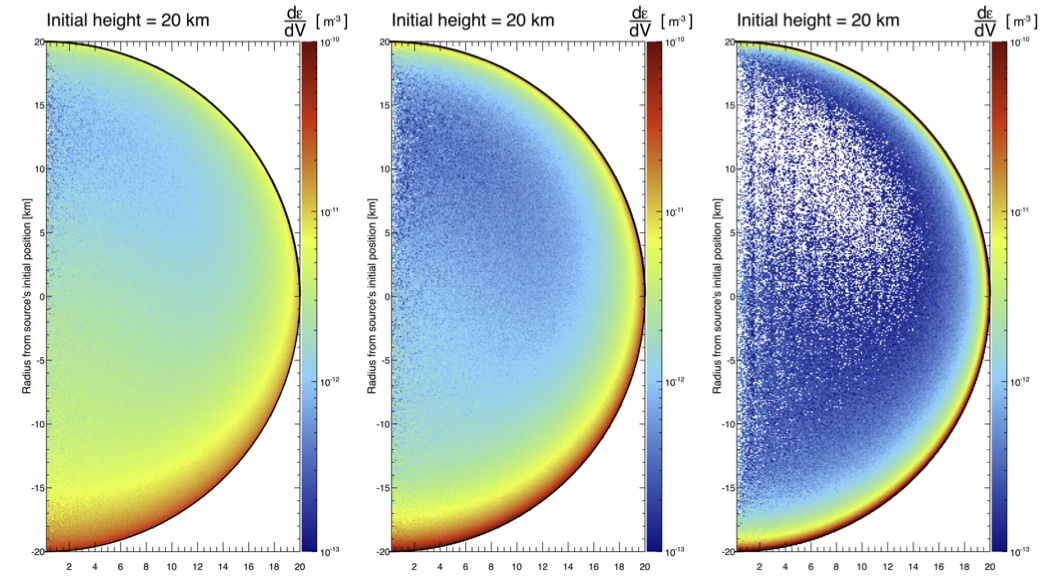}
\caption{(Color online) {\bf Effect of changing the $\mathbf{g}$ value on the density distribution of photons propagating from an isotropic source.} Simulations are for atmospheres of only aerosols with \lambdaaerzero$=14.2$~km and \haerzero$=8$~km {\it i.e.}\ an aerosol density distribution similar to molecules. Results are presented for three different values of $g=\{0.3,0.6,0.9\}$, left, middle and right, respectively.}
\label{fig:changephasefunction} 
\end{figure*}

\section{Distribution of scattered photons from an isotropic source in the atmosphere}
\label{sec:simu_results}
The Monte Carlo simulation is used to study the distribution of scattered photons across space for an isotropic light source. The variable $\varepsilon$ is introduced as the fraction of total energy of the isotropic source in a given histogram bin. In this simulation which is for a fixed wavelength ($\lambda = 350~$nm), this is given by the number of photons in a bin divided by the total number of photons $N$. Then, the density per unit volume of the fraction of initial energy $\mathrm{d}\varepsilon/\mathrm{d}V(r_{\mathrm{rel}}, \theta_{\mathrm{rel}})$  is calculated by dividing by the elemental volume of each bin. The elemental volume d$V$ of each bin is equal to $\mathrm{d}V = 2\pi r_{\mathrm{rel}}^{2}\sin(\theta_{\mathrm{rel}})\mathrm{d}r_{\mathrm{rel}}\mathrm{d}\theta_{\mathrm{rel}}$, where $\mathrm{d}r_{\mathrm{rel}}$ and $\mathrm{d}\theta_{\mathrm{rel}}$ are the widths of each bin in \rrel and \thetarel, respectively. 
The quantity $\mathrm{d}\varepsilon/\mathrm{d}V(r_{\mathrm{rel}}, \theta_{\mathrm{rel}})$ is such that the total fraction of energy in a given volume $\varepsilon_{\mathrm{total}}$ is given by 
\begin{equation}
\varepsilon_{\mathrm{total}} = \iiint \frac{\mathrm{d}\varepsilon}{\mathrm{d}V} r_{\mathrm{rel}}^{2}\sin(\theta_{\mathrm{rel}})\mathrm{d}r_{\mathrm{rel}}\mathrm{d}\theta_{\mathrm{rel}}\mathrm{d}\phi_{\mathrm{rel}} ,
\end{equation}
where limits of the integral are chosen to represent this volume. Figure~\ref{fig:relativedistributions} displays the density of indirect photons (scattered photons) for an isotropic light source at an initial height of $20~$km after a distance of propagation of $20~$km. On each plot, a black semicircle with radius \rdir is drawn to represent the position of direct (unscattered) photons. At $\theta_{\mathrm{rel}}=0\degree$ ({\it i.e.}\ positive vertical axis), \rrel~extends in a direction directly above the initial source's position and at $\theta_{\mathrm{rel}}=180\degree$ ({\it i.e.}\ negative vertical axis) directly to the ground. Aerosols are less dense than molecules in a real atmosphere and the object of this section is to show that this difference in density means aerosols have a very small effect on the overall distribution of scattered photons across space. The multiplicative scale factor for aerosols is set to \lambdaaerzero$=10$~km to represent a density of aerosols that is higher than likely to be found in a real atmosphere. Simulations are run independently for atmospheres of only molecules (left), only aerosols (middle) and both being simultaneously present (right). It is directly evident by eye from the striking similarity between Fig.~\ref{fig:relativedistributions}(left) and Fig.~\ref{fig:relativedistributions}(right)  that the overall distribution of indirect photons in the atmosphere is governed by molecules. In spite of the negligible effect of aerosols on the overall distribution of indirect photons, their presence should not be forgotten, in particular near to ground level where the ground-based detectors are located.

\begin{figure*}[p]
\centering
\includegraphics [width=1.0\textwidth] {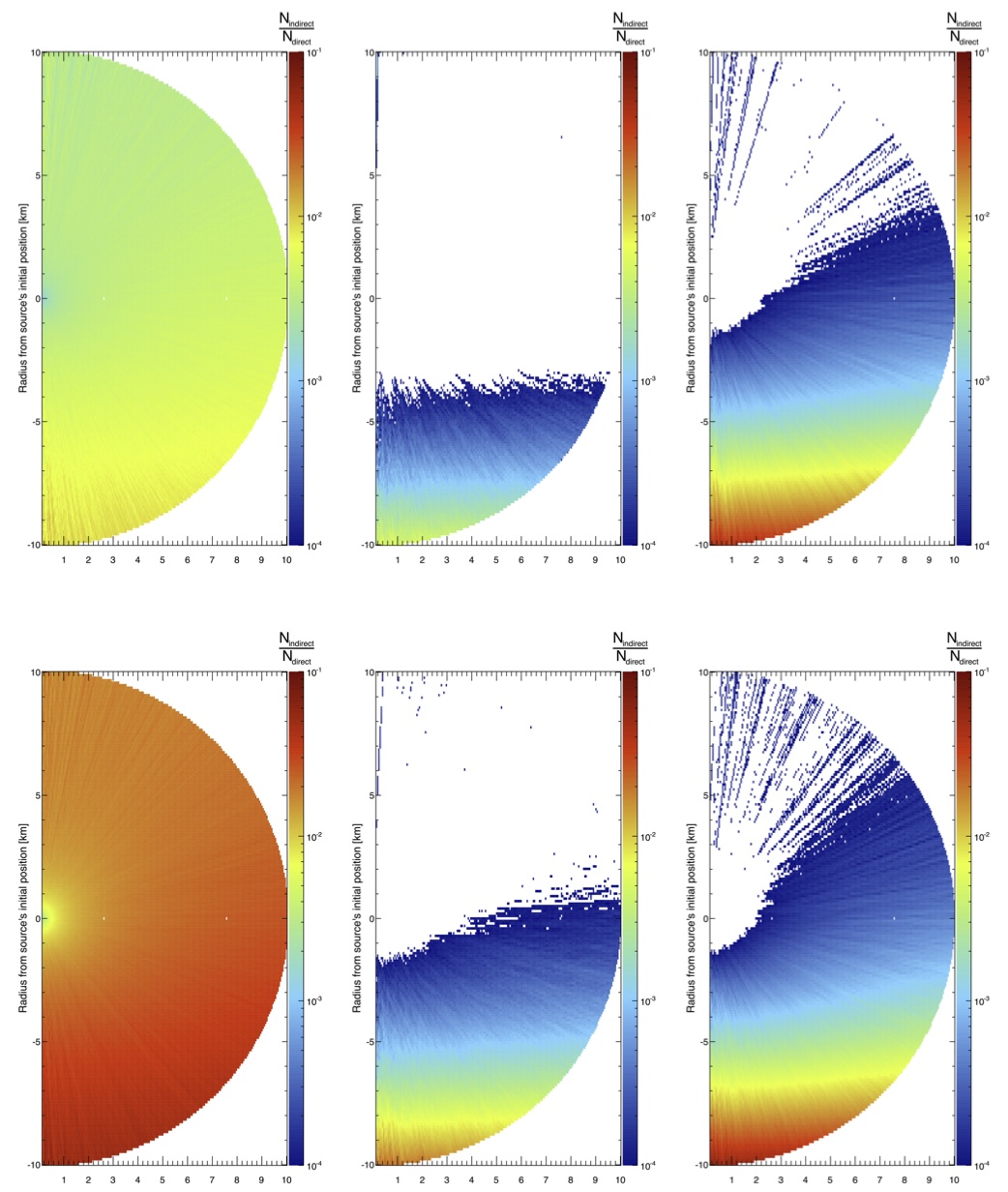}
\caption{(Color online) {\bf Effect of the scattering phase function and the detection time}. Simulations are run for sources of initial height $h_{\mathrm{init}}=10$~km and a detection time of $t_{\mathrm{det}}=100$~ns for the top panel and $t_{\mathrm{det}}=1000$~ns for the bottom panel, individually for atmospheres of molecules only (left) or aerosols only, $g=0.3$ (middle) or $g=0.9$ (right). Aerosol density parameters are \{\lambdaaerzero$=25.0$~km, \haerzero$=1.5$~km\}. Ray structures are related to a lack of statistics for simulation of scattered photons.}
\label{fig:tdec} 
\end{figure*}

This part demonstrates how changing the dominating scattering phase function and, in particular varying $g$, changes the dispersion of scattered photons across space and time. To make evident the effects, simulations are run independently for atmospheres of only aerosols and molecules. Atmospheres of only aerosols are given an unrealistic density distribution set to be the same as molecules {\it i.e.}  \{\lambdaaerzero$=14.2$~km, \haerzero$=8.0$~km\}. The initial height is 20~km for all isotropic sources and the distance propagated is $D=20$~km. Figure~\ref{fig:changephasefunction} shows the results for atmospheres of only aerosols with three different values of $g=\{0.3,0.6,0.9\}$ whilst Fig.~\ref{fig:relativedistributions}(left) shows the equivalent result for an atmosphere consisting of molecules only. In Fig.~\ref{fig:relativedistributions}(left) all scattering events are governed by the molecular phase function. This gives a fairly isotropic distribution of the direction of scattered photons across space. There is no notable accumulation of indirect photons in any given area other than slightly before \rdir. Contrastly, in Fig.~\ref{fig:changephasefunction}, the density distribution of indirect photons across space is much more anisotropic. The important point made evident through this figure is that as $g$ increases, more scattered photons are found close to \rdir. Moreover, since the aerosol density distribution is set to the same values than the molecular component in Fig.~\ref{fig:relativedistributions}(left), this effect can be purely accredited to the changing scattering phase functions. The explanation of this trend lies in the increasing anisotropy of the directions of scattered photons for increasing $g$. For a photon scattered through a scattering angle $\psi$, its component of direction along the direction of direct light is $\cos\psi$. As such, for lower values of $\psi$, the component of direction along the direction of direct light is higher. Figure~\ref{eq:APF} clearly shows that, for higher values of $g$, lower scattering angles of $\psi$ are more likely to be generated and hence explains the increased accumulation of indirect photons just before \rdir. The next section is devoted to demonstrate that, even in much smaller proportions than molecules, the high forward scattering peak associated with aerosol scattering events is important in considering the indirect light signal recorded by ground-based detectors.

\section{Global view of indirect photon contribution to the total light detected}
\label{sec:global_view}
This section aims to observe how different aerosol conditions, and especially different scattering phase functions, affect the ratio of indirect to direct light arriving at detectors within a given time interval (or integration time) $t_{\mathrm{det}}$ across all space. The simulation is used to propagate photons from an isotropic source for a given distance $D$, at which point, values of position are stored for direct photons only. Indirect photons are then simulated to propagate for a further amount of time $t_{\mathrm{det}}$. Any of these indirect photons crossing the sphere of direct photons with radius $D$ within the time $t_{\mathrm{det}}$ are considered detected. With respect to the position in space that each histogram bin holds data for, the histograms presented in this section have the same format as in Section~\ref{sec:simu_results}. However, in this section, each histogram bin now represents the ratio of indirect to direct photons $N_{\mathrm{indirect}}/N_{\mathrm{direct}}$ detected at the point \{\rrel, \thetarel \}, within the interval of time starting when direct photons reach the point and finishing within a time $t_{\mathrm{det}}$ later.

Simulations here are run separately for atmospheres of only molecules or aerosols. Density parameters of \{\lambdaaerzero=25.0~km, \haerzero=1.5~km\} for the aerosol population are deliberately chosen such that the effects observed can not be simply accredited to an over-estimated density of aerosols in the atmosphere. Figure~\ref{fig:tdec}(top) shows results for a detection time of $t_{\mathrm{det}}=100$~ns for atmospheres of molecules only (left) and aerosols only with values of $g=\{0.3, 0.9\}$ (middle and right, respectively). For all configurations, there is an increasing ratio of indirect to direct photons observed towards ground level. This is expected as the amount of direct photons decreases and indirect photons increases for the increasing concentration of scatterers at lower heights. Of much greater interest is the fact that at ground level, $N_{\mathrm{indirect}}/N_{\mathrm{direct}}$ for aerosols with a high $g$ value is much greater than $N_{\mathrm{indirect}}/N_{\mathrm{direct}}$ for molecules (in spite of a much lower concentration). This directly demonstrates that, for low detection times, a high value of $g$ has an influence on the ratio $N_{\mathrm{indirect}}/N_{\mathrm{direct}}$ that outweighs the fact that aerosols are at a lower density than molecules. It can be explained by referring back to Fig.~\ref{fig:changephasefunction}, where an increasing value of $g$ leads to an increasing accumulation of indirect photons just before the direct photon ring. However, this amount of scattering by aerosols begins to become significant enough only at low heights above ground level. This is an important fact for ground-based detectors.

Turning attention now to Fig.~\ref{fig:tdec}(bottom), the same results are shown for $t_{\mathrm{det}}=1000$~ns. Looking at Fig.~\ref{fig:tdec}(top, right) and Fig.~\ref{fig:tdec}(bottom, right), it is evident that the change in the ratio $N_{\mathrm{indirect}}/N_{\mathrm{direct}}$ is nearly invisible when increasing $t_{\mathrm{det}}$ from 100 to 1000~ns for $g=0.9$. This implies that for a very high $g$ value, the total amount of indirect photons that will ever be detected are nearly all detected at a very low detection time. Figure~\ref{fig:tdec}(bottom, left) equally shows that molecules begin to have a more prominent effect than aerosols for greater detection times. Indeed, in the case of a higher detection time, a photon being much further from the direct photon sphere has enough time to reach the sphere and be detected. In contrast, for a lower detection time, a high forward scattering peak is necessary for the photons to be close enough to the direct photons and arrive within this detection time. It is therefore the relative density of the scatterers in the atmosphere that bares more influence on the ratio $N_{\mathrm{indirect}}/N_{\mathrm{direct}}$ for higher values of $t_{\mathrm{det}}$. Hence, taking into account the relative density of molecules and aerosols in the atmosphere, the multiple scattering caused by aerosols is not negligible near to ground level, especially for large values of asymmetry parameter $g$ and low detection times $t_{\mathrm{det}}$. The next section continues to investigate the effect of changing $g$ and $t_{\mathrm{det}}$ but for the specific case of a ground-based detector.

\begin{figure}[b!]
\centering
\includegraphics[width =1.05\columnwidth]{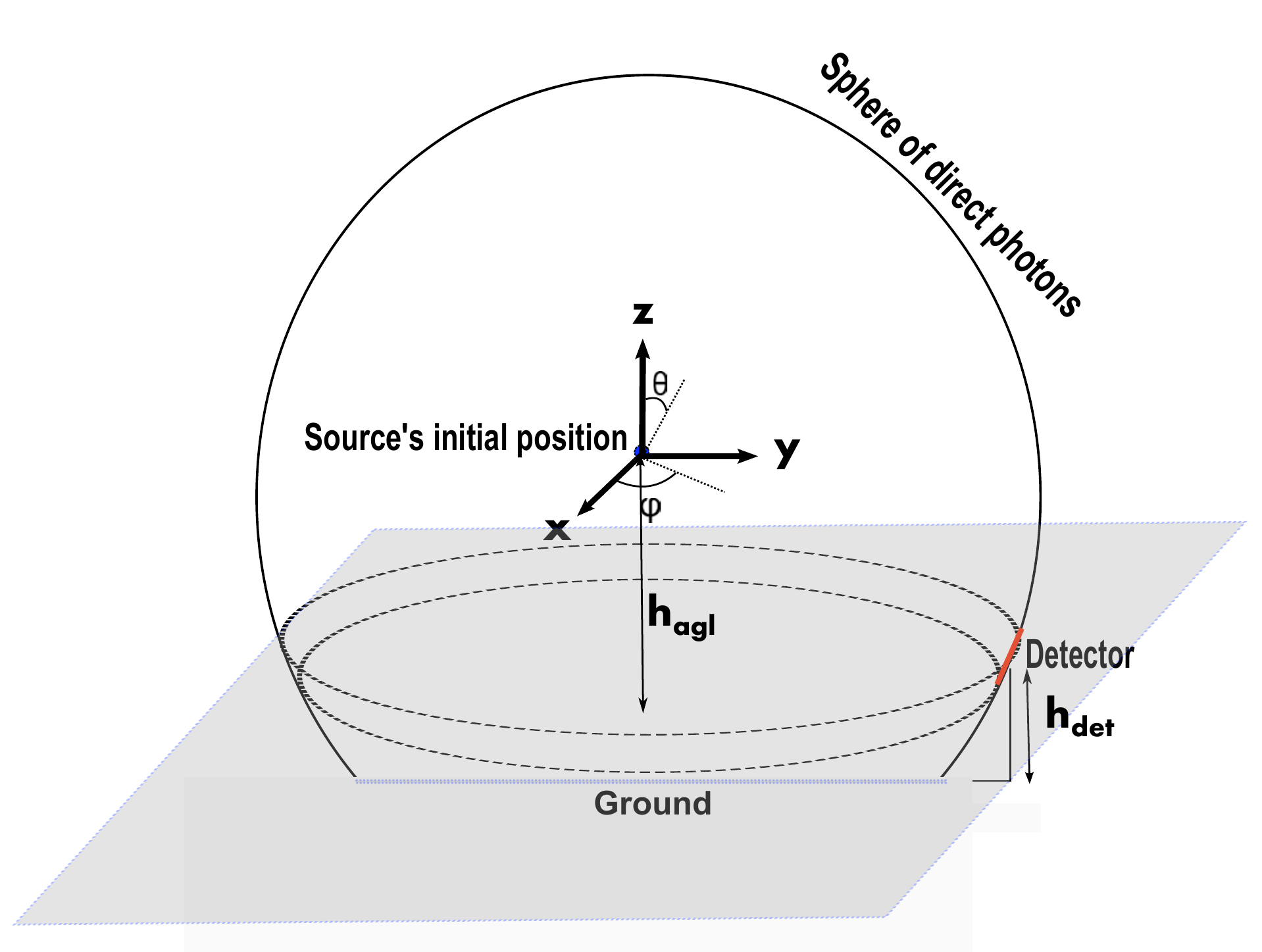}
\caption{A diagram showing how the detector is simulated to have an extent of $2\pi$ in azimuthal angle to increase the amount of statistics retrieved for indirect photons.}
\label{fig:det}
\end{figure}

\begin{figure*}[t!]
\includegraphics [width=1.0\textwidth] {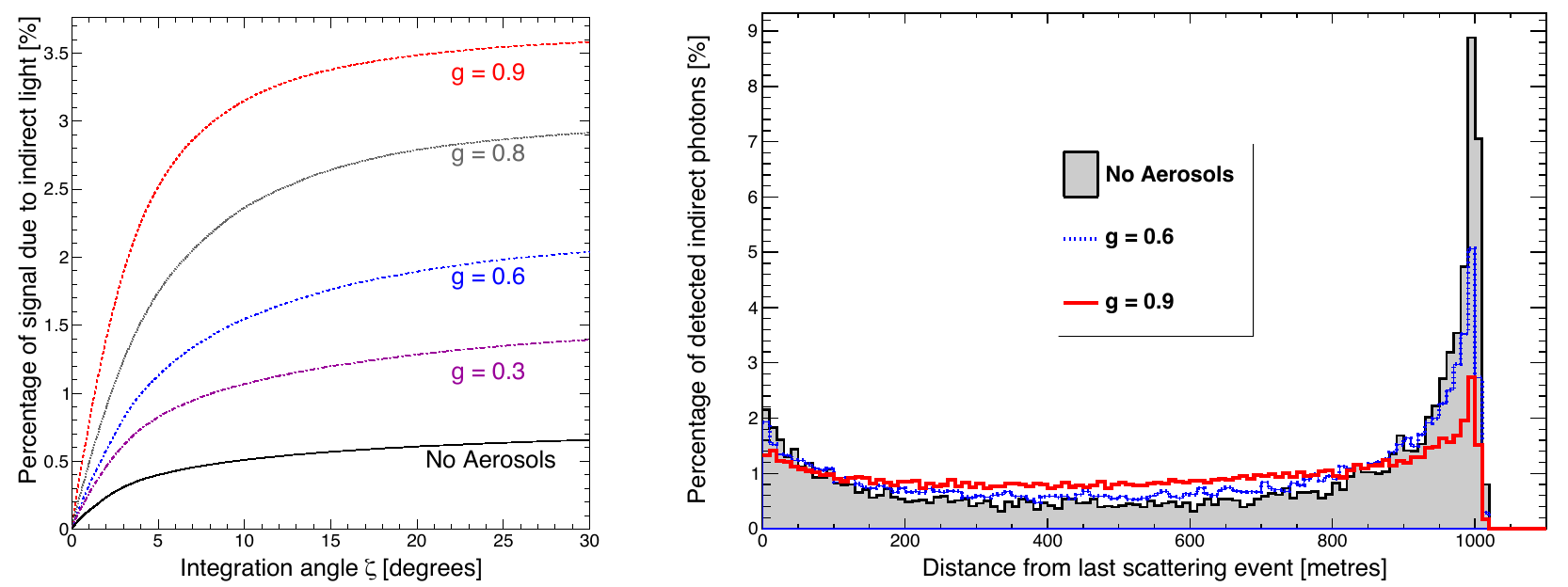}
\caption{(Color online) {\bf Plots for an isotropic source placed at $\theta_{\rm inc}=3\degree$, $D=1$~km for atmospheres where aerosols and molecules are simultaneously present.} The aerosol concentration and the detection time are kept constant at \lambdaaerzero= 25~km and $t_{\rm det} = 100~$ns, respectively. (left) Percentage of signal due to indirect light for different integration angles $\zeta$ and different $g$ values. The black line is an exclusive case where only molecules are present. (right) Percentage of the detected indirect photons that were last scattered at different distances from the detector for the three same cases.}
\label{fig:PSF} 
\end{figure*}

\section{Atmospheric point spread function for a ground-based detector}
\label{sec:ground_telescope}
The quantity of multiple scattered light recorded by an imaging system or telescope is of principal interest, and especially this contribution as a function of the integration angle $\zeta$. The angle $\zeta$ is defined as the angular deviation in the entry of indirect photons at the detector aperture with respect to direct photons. For direct photons from an isotropic source, the angle of entry is usually approximated to be constant as the entry aperture of the detector is always very small relative to the distance of the isotropic sources. In contrast, multiply scattered photons can enter the aperture of the detector at any deviated angle $\zeta$ from the direct light between 0\degree~and 90\degree. The value $\zeta$ for each indirect photon entering the detector is calculated by considering its deviation from direct light in elevation and azimuthal angle noted $\Delta\theta$ and $\Delta\phi$, respectively: $\zeta = \cos^{-1} [ \cos\Delta\theta \cos\Delta\phi]$. A Taylor expansion of this equation, keeping all terms up to second order, means that $\zeta$ can approximately be written as $\zeta \approx \sqrt{\Delta\theta^2 + \Delta\phi^2}$.

The main problem in simulating indirect light contribution at detectors is obtaining reasonable statistics within reasonable simulation running times. The root of the problem is the very small surface area of the detector relative to the large distances where isotropic sources are created. The amount of direct photons is calculated analytically by modelling the detector as a point relative to the initial position of the isotropic source. Thus, all direct photons are considered to follow the same path and the infinitesimal change in the number of direct photons d$N_{\mathrm{direct}}$ for an infinitesimal step length d$l$ is then written as $\mathrm{d}N_{\mathrm{direct}}= - \left[{N_{\mathrm{direct}} \,\mathrm{d}l}/{\Lambda_{\mathrm{mol}}(l)} + {N_{\mathrm{direct}} \,\mathrm{d}l}/{\Lambda_{\mathrm{aer}}(l)}\right]$. The same approach as explained in~\cite{MS_roberts} is used to cut running times of the simulation for indirect photons. The symmetry in the distribution of scatterers in azimuthal angle, as explained in Section~\ref{sec:phase_functions}, is once again applied here. This symmetry means that so long as the detector has the same height and distance from the source, the azimuthal angle relative to it is unimportant. As such, the surface area of the detector is increased in the simulation by extending it through an azimuthal angle of $2\pi$ so that a greater amount of indirect photons is detected and better statistics are obtained. The setup of the extended detector is drawn in Fig.~\ref{fig:det}, where the strip of the sphere has a width corresponding to the diameter of the detector. Also, stopping the tracking of all photons that can no longer be detected further reduces the simulation time.

\begin{figure*}[!t]
\centering
\includegraphics [width=1.0\textwidth] {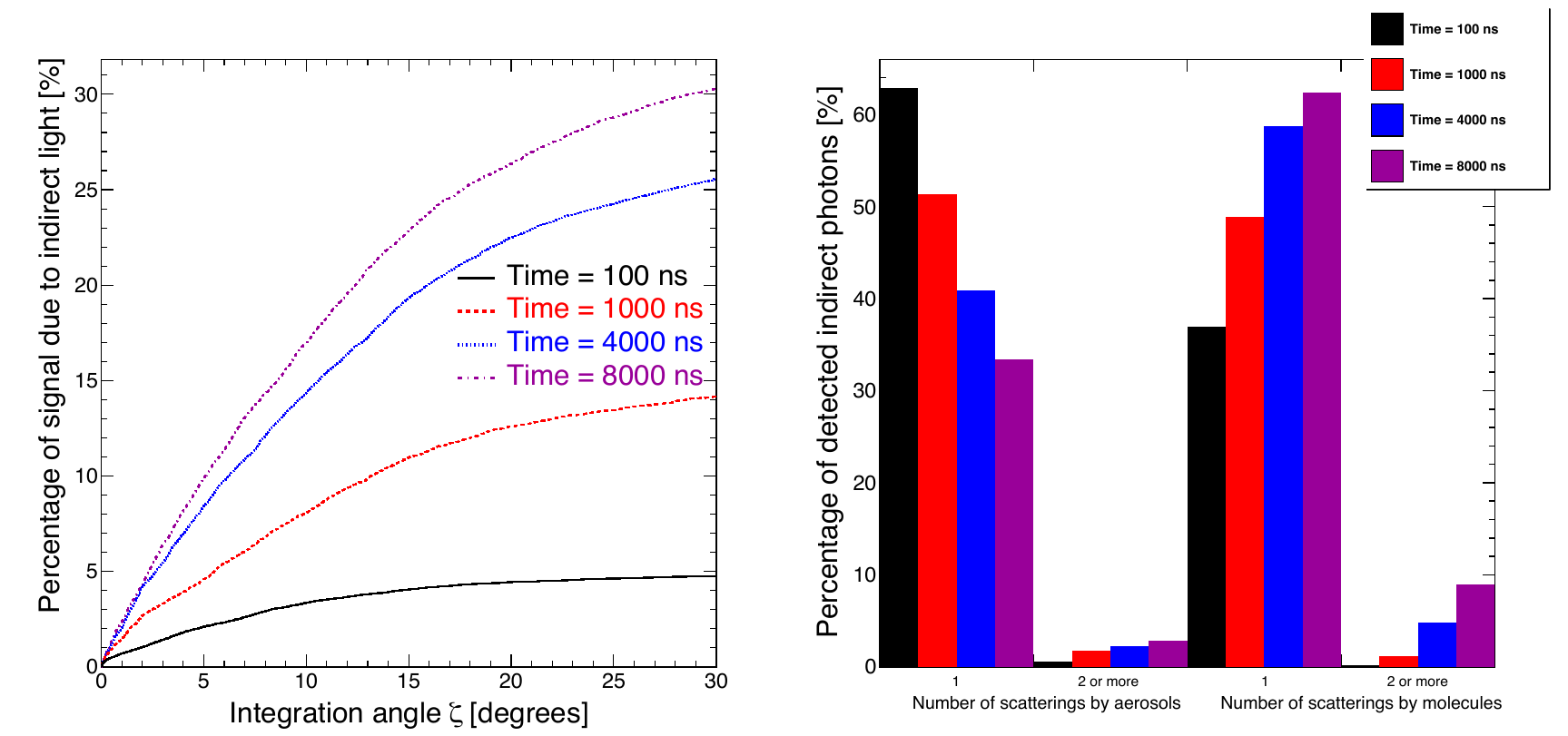}
\caption{(Color online) {\bf Plots to demonstrate the effect of increasing integration time $t_{\rm det}$.  The inclination angle is $\theta_{\mathrm{inc}}=15\degree$ and the  distance $D=30$~km. \lambdaaerzero= 25~km and $g=0.6$.} (left) Percentage of signal due to indirect light for different integration angles $\zeta$ and different integration times $t_{\mathrm{det}}$. (right) Percentage of detected indirect photons that have undergone either 1 or 2 and more scatterings by aerosols or molecules.}
\label{fig:timechange}
\end{figure*}
 
This part aims to look at the effect of changing aerosol size (via the asymmetry parameter $g$) on the amount of indirect light recorded at the detector for an isotropic source at different positions. The integration time of the detector is set to $t_{\mathrm{det}} = 100$~ns and the aerosol attenuation length is fixed at \lambdaaerzero=25~km. The percentage of light due to indirect photons against integration angle $\zeta$ is given by the ratio (indirect light) over (direct light + indirect light), where the direct or indirect light signals are the number of photons collected within the given integration angle $\zeta$. Figure~\ref{fig:PSF}(left) shows the results for an isotropic source placed at a distance of $D=1$~km and at a very low inclination angle of $\theta_{\mathrm{inc}}= 3\degree$ ($\theta_{\mathrm{inc}} = \sin^{-1}(h_{\rm source}/D)$, where $h_{\rm source}$ is the height of the source above ground level). As expected, the amount of signal due to indirect light increases consistently with integration angle $\zeta$ as all direct light arrived at $\zeta=0\degree$. These curves are directly linked to the point spread function since only a differentiation with respect to $\zeta$ is needed. These curves are similar to measurements done, for instance, by Bissonnette~\cite{Bissonnette,BenDor}. A more interesting feature is the increased contribution from indirect light for increasing aerosol size ({\it i.e.}\ a higher value of the asymmetry parameter $g$). Comparing the situation of an atmosphere with no aerosols to one with aerosols, atmospheric aerosols have a non-negligible effect on the percentage of indirect light received at a ground-based detector. The explanation of this observation lies once again in the anisotropy associated with the scattering phase function. The percentage of indirect photons last scattered at a given distance relative to the detector is plotted in Fig.~\ref{fig:PSF}(right). In the case of an atmosphere with only molecules, most of scattered photons recorded within the integration time of $100~$ns originate from the position of the isotropic source. Indeed, there is an increased density of photons at this distance where the source is initiated. For the case of larger aerosol sizes, corresponding to larger $g$ values, the peak at the source's initial position is now greatly diminished. This is a result of detection of photons scattered more uniformly across all distances. This occurs because aerosols with higher values of $g$ have a higher probability of scattered photons in a very forward direction.

Figure~\ref{fig:timechange}(left) displays results of simulations run for the same geometrical configuration and aerosol parameters but different integration times. The distance to the detector is set to be $D=15$~km, inclination angle $\theta_{\mathrm{inc}}=15\degree$, $g= 0.6$ and \lambdaaerzero= 25~km. It is observed that for an increasing time, the total amount of signal due to indirect light also increases. For the 900~ns interval between $t_{\mathrm{det}}=100$~ns and $t_{\mathrm{det}}=1000$~ns, the amount of signal due to indirect light increases greatly, meaning that  there are still a lot of indirect photons yet to arrive after the integration time of 100~ns. However, this observation is not true anymore after a large time delay, typically greater than $4000~$ns in our case. The bar chart displayed in Fig.~\ref{fig:timechange}(right) shows the percentage of detected indirect photons having undergone 1 or 2 and more scatterings by aerosols or molecules. The most notable feature of this chart is the increasing amount of molecularly scattered photons detected for increasing integration times. As previously explained in Section~\ref{sec:global_view}, aerosols begin to play a lesser role for longer detection times. This is because photons that were scattered at  positions that are not necessarily close to the path of direct light, now have enough time to travel and reach the detector within this longer integration time. This idea is confirmed by an increased uniformity across space in the distance from the last scattering event for increasing integration times (plot not shown here).

\section{Conclusion}
A new Monte Carlo simulation for the scattering of light has been created and used to observe atmospheric aerosol effects on the percentage of indirect light collected by detectors. The study began with a general description of the dispersion of scattered photons in different atmospheric conditions. It was found that for an increased value of the asymmetry parameter $g$ ({\it i.e.}\ a larger aerosol size), a greater accumulation of scattered photons close to the direct photons is found. The principal argument presented in this work is that, even for a low density of aerosols in the atmosphere, the ratios of indirect to direct photons detected can be comparable or greater to those caused by molecules. In particular, the value of detection time $t_{\mathrm{det}}$ is proved to play an important role on the relative effects of molecules and aerosols on the ratio. This phenomenon is also used to estimate the aerosol size distribution, especially for very large aerosols. The technique is described in detail in~\cite{Zacanti,Trakhovsky_1,Trakhovsky_2}.

In addition, this aerosol size effect could still solve some unsolved experimental observations as the measurement done at the Pierre Auger Observatory~\cite{PAO,MyICRC} a few years ago. Indeed, part of the point spread function measured by ground-based telescopes is still not fully understood, {\it i.e.}\ cannot be reproduced in simulations~\cite{ICRC_Julia,MS_Assis}. One of possible explanations could be an additional contribution coming from a population of large aerosols present in the atmosphere.

\section*{Acknowledgements}
One of the authors, KL, thanks Marcel Urban for having been at the beginning of this study. Also, the authors thank their colleagues from the Pierre Auger Collaboration for fruitful discussions and for their comments on this work.

\end{document}